\documentclass[twocolumn,amssymb, nobibnotes, showpacs, superscriptaddress, aps, prd]{revtex4}
\usepackage{graphicx,amsmath,amssymb,color}

\begin{document}
\title{Symmetry-Breaking Zeeman-Coherence Parametric Wave Mixing Magnetometry}

\author{Feng Zhou}
\affiliation{National Institute of Standards \& Technology, Gaithersburg, Maryland USA 20899}
\author{Chengjie Zhu}
\affiliation{School of Physical and Engineering Sciences, Tongji University, Shanghai, China 200092}
\author{E. W. Hagley}
\affiliation{National Institute of Standards \& Technology, Gaithersburg, Maryland USA 20899}
\author{L. Deng}
\affiliation{National Institute of Standards \& Technology, Gaithersburg, Maryland USA 20899}

\date{\today}

\begin{abstract}
\noindent {\bf The nonlinear magneto-optical effect has significantly impacted modern society with prolific applications ranging from precision mapping of the Earth's magnetic field to bio-magnetic sensing. Pioneering works on collisional spin-exchange effects have led to ultra-high magnetic field detection sensitivities at the level of $fT/\sqrt{Hz}$ using a single linearly-polarized probe light field. Here we demonstrate a nonlinear Zeeman-coherence parametric wave-mixing optical-atomic magnetometer using room temperature rubidium vapor that results in more than a three-order-of-magnitude optical signal-to-noise ratio (SNR) enhancement for extremely weak magnetic field sensing. This unprecedented enhancement was achieved with nearly a two-order-of-magnitude reduction in laser power while preserving the sensitivity of the widely-used single-probe beam optical-atomic magnetometry method. This new method opens a myriad of applications ranging from bio-magnetic imaging to precision measurement of the magnetic properties of subatomic particles.}
\end{abstract}

\pacs{42.65.-k, 07.55.Ge, 78.20.Ls, 32.60.+i}
\maketitle

\vskip 5pt
\noindent{\bf Significance}
\vskip 5pt
\noindent{\bf We report nearly three orders of magnitude increase in the magnetic field sensing signal-to-noise ratio using a novel light-matter interaction scheme.  This unprecedented advance in precision magnetic field measurement is achieved simultaneously with significant reduction in laser power while perfectly preserving magnetic field sensitivity. }

\vskip 20pt
\noindent The fundamental physical process behind the nonlinear Faraday rotation effect \cite{book1} is the generation of a Zeeman coherence by a linearly polarized probe field coupling different Zeeman-shifted sub-levels of an atom in the presence of a magnetic field \cite{BRreview,book2}. In the semi-classical picture different polarization components of a linearly polarized light field experience different refractive indexes when the medium is subject to a magnetic field, and this gives rise to an optical polarization rotation detectable using standard polarimetry methods. For a three-state atomic system coupling to a linearly-polarized probe field in a $\Lambda-$configuration \cite{BRreview,Tslope,note3e,note4} (Fig. 1a), assuming that the initial population is equally distributed among the two Zeeman-ground states, the nonlinear magneto-optical polarization rotation arising from the differential phase shift of the two circular polarization components comprising the linearly-polarized probe light can be derived using a third-order perturbation theory \cite{shen}:

\begin{subequations}   
\begin{eqnarray}
P_{\pm}^{(3)}
&=&\mp\frac{2\kappa\Gamma\rho_{11}^{(0)}\Omega_p^{(\pm)}}{(\delta_p+i\Gamma)( 2\delta_B\pm i\gamma_0)}\left(\frac{|\Omega_p^{(\mp)}|^2}{\delta_p^2}\right),\\
\frac{\partial\Omega_{\pm}}{\partial z}&=&P_{\pm}^{(3)}(\omega_p).
\end{eqnarray}
\end{subequations}

\begin{figure}
  \centering
  \includegraphics[width=8.5 cm,angle=0]{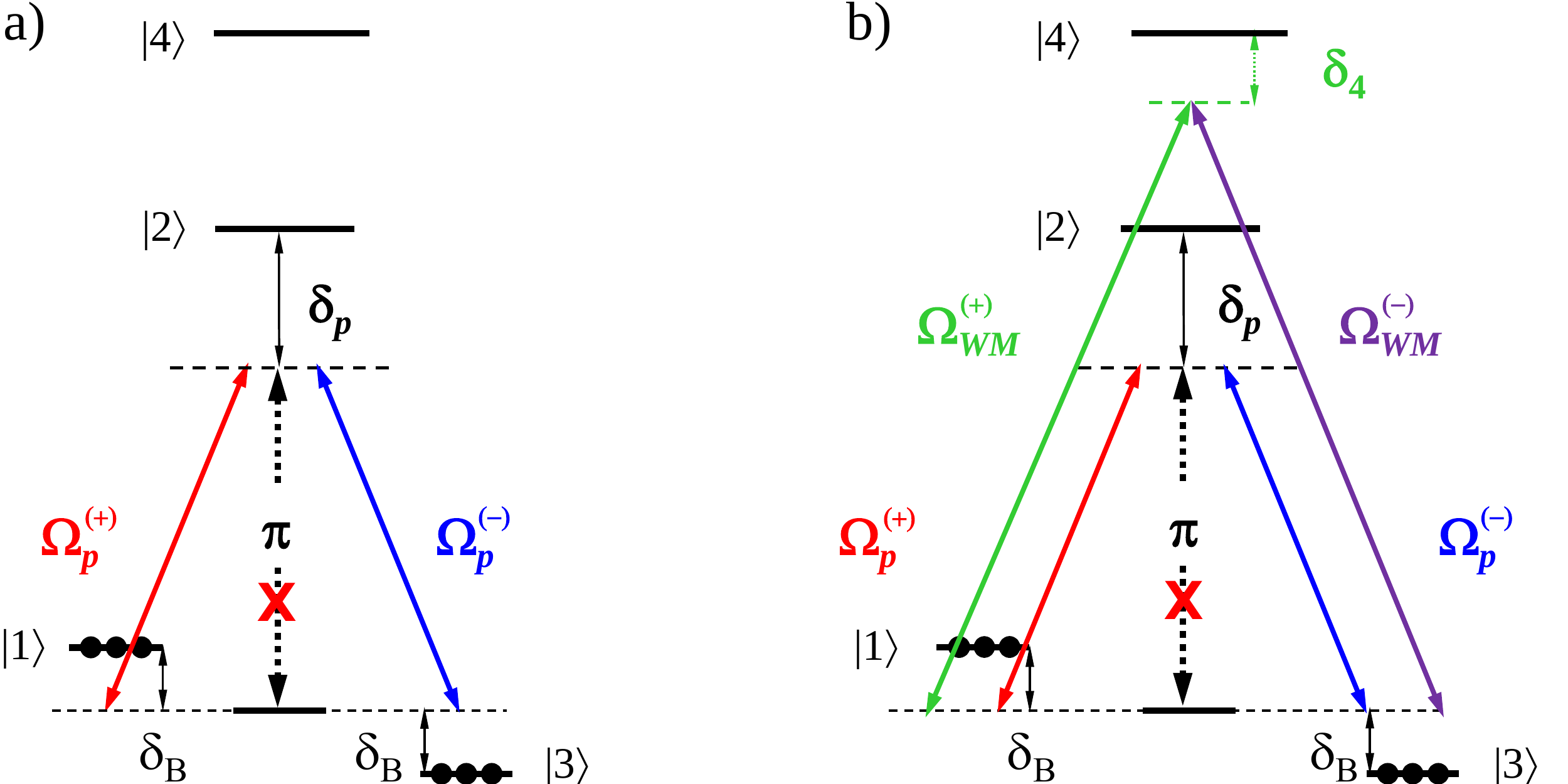}
  \caption{(a): Single-probe $\Lambda-$scheme in the presence of a magnetic field. The lower three states are the magnetic sub-levels of the $F=1$ manifold (from left  $m_F=-1,0,+1$). (b): Symmetry-breaking Zeeman-coherence parametric wave-mixing (WM) scheme with linearly-polarized WM light field. For mathematical simplicity only excited states $|2\rangle$ and $|4\rangle$ are considered. }
\end{figure}

\noindent In Eq. (1a) $\delta_B=\mu_Bg_FB_z$ is the magnitude of the magnetic field ($B_z$) induced Zeeman shift, and this familiar result is applicable when power broadening and two-photon saturation effects are negligible.  
$\kappa$ is proportional to the product of the medium density and the relevant optical transition strength. $\rho_{ij}$ represents the coherence ($i\ne j$) and population ($i=j$), and we have taken $\rho_{11}^{(0)}=\rho_{33}^{(0)}$. $\Omega_p^{(\pm)}=D_{ij}E_p^{(\pm)}/\hbar$ are the Rabi frequencies of the circular components of the linearly-polarized probe field $\vec{E}_p=\vec{E}_p^{(+)}+\vec{E}_p^{(-)}$, where $D_{ij}$ is the dipole matrix element of the relevant transition. 
$\delta_p$ is the one-photon probe laser detuning from the excited electronic state having a resonance line width of $\Gamma=\gamma_{21}=\gamma_{23}$. Equation (1a) has a Lorentzian line shape with a Zeeman resonance line-width $\gamma_0$ (Zeeman de-coherence rates of the corresponding magnetic sub-levels are taken to be $\gamma_0=\gamma_{31}=\gamma_{13}$). 
Seminal studies on collisional spin-exchange and relaxation effects \cite{happer1,happer2,walker,romalis1,romalis2,romalis3,budker1,budker1a,russian1,russian2,kitching,ref17a} have substantially reduced the Zeeman de-coherence rate $\gamma_0$, enabling ultra-high magnetic field sensitivities \cite{romalis2,budker1a}.

\begin{figure}
  \centering
  \includegraphics[width=8.5 cm,angle=0]{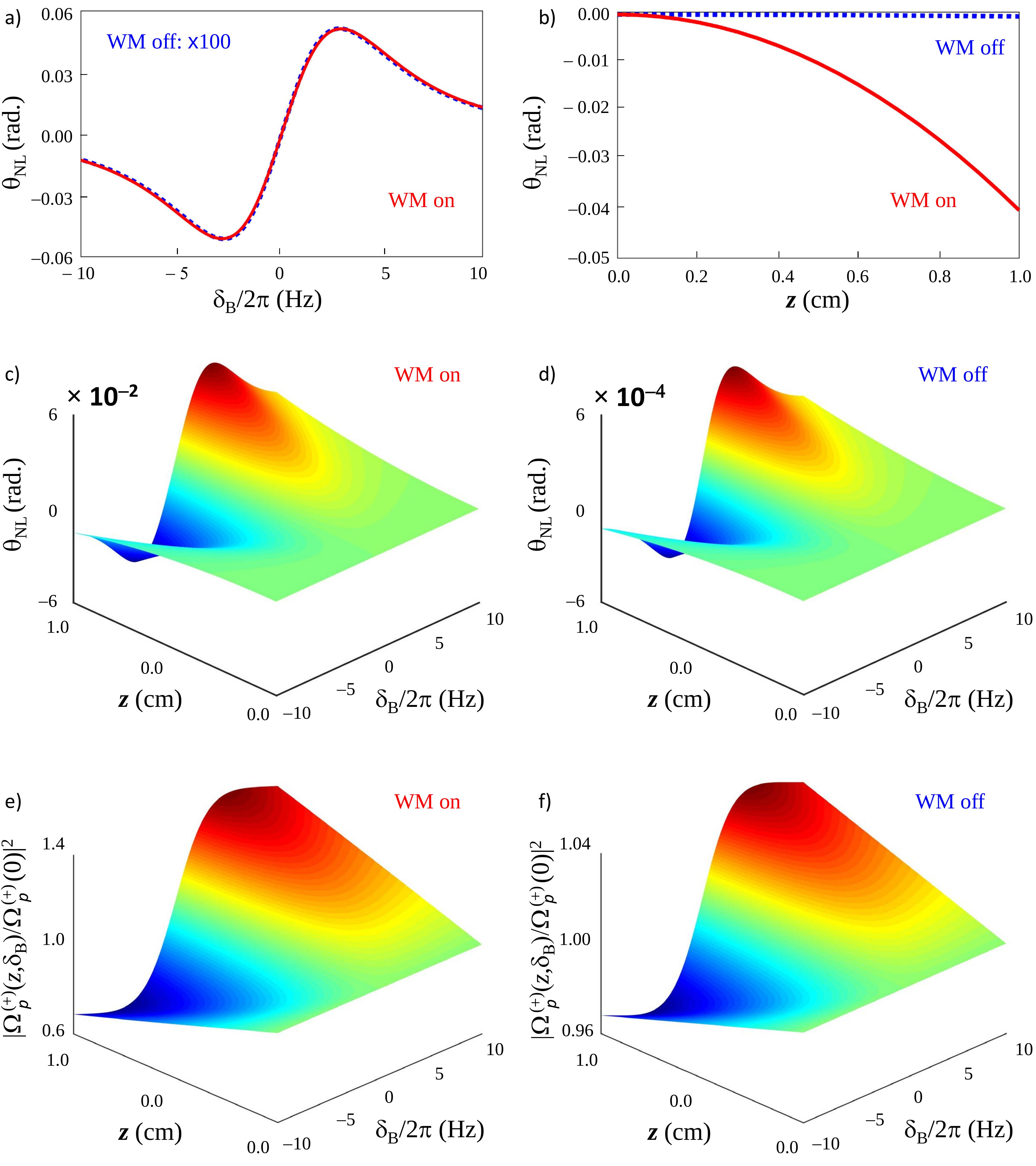}
  \caption{(a) Nonlinear polarization rotation of the single-probe $\Lambda-$scheme (dotted blue) and the Zeeman-coherence parametric WM scheme (solid red) as a function of magnetic detuning $\delta_B$ at $z$ = 1 cm.  (b) Nonlinear polarization rotation as a function of propagation distance $z$ for $\delta_B$ = $-$5 Hz. The superior performance of parametric WM scheme (solid red) cannot be matched by simply increasing the intensity of the single-probe beam scheme (dotted blue). (c) and (d): Nonlinear polarization rotation as a function of $z$ and $\delta_B$ with (c) and without (d) the WM field. (e) and (f): Normalized $\Omega_p^{(+)}$ component of the probe field as a function of $z$ and $\delta_B$ with (e) and without (f) the WM field. Figure (f) clearly exhibits the self-limiting effect imposed by symmetry, as shown in Eq. (2). Parameters: $\Omega_p^{(\pm)}(z=0)/2\pi$ = 200 kHz, $\Omega_{WM}(z=0)/2\pi$ = 300 kHz, $\delta_p/2\pi$ = 1 GHz, $\delta_{4}/2\pi$= 0.5 GHz, $\Gamma/2\pi$ = 10 MHz, $\gamma_0/2\pi$ = 10 Hz, $\kappa$ = 10$^9$/(cm.s), $\rho_{11}^{(0)}=\rho_{33}^{(0)}$ = 0.5. }
\end{figure}
\vskip 5pt
\noindent The single probe beam $\Lambda-$scheme is elegantly simple with two symmetric channels where the identical transition strengths result in canceled residual frequency shifts.  In this case a significant Zeeman coherence \cite{note3} can be established by the two circular components of the probe field in a two-photon resonant transition where each component simultaneously undergoes stimulated emission and absorption. However, because the probe field is the only source of energy, the symmetric nature of the process dictates that both components cannot grow and the scheme is self-limiting by detailed balance. Consequently, we have 

\begin{equation}
\frac{\partial |\Omega_{\pm}|}{\partial z}\simeq 0,\quad {\rm and}\quad \theta_{NL}\propto\; nL,
\end{equation}
where $n$ and $L$ are the medium density and length, respectively, and $\theta_{NL}$ is the nonlinear polarization rotation. 
  
\vskip 5pt
\noindent
The Zeeman-coherence parametric wave-mixing scheme shown in Fig. 1b breaks the self-limiting symmetry of the probe field by introducing a wave-mixing (WM) process which enables efficient energy transfer from a second light field $\vec{E}_{WM}$ to the selected branch of the probe field through coherently populated intermediate states with a large Zeeman coherence \cite{lutwm1,lutwm2}.  For the different probe components this parametric WM process yields \cite{shen}

\begin{subequations}
\begin{eqnarray}
P_{\pm;WM}^{(3)}&\!=\!&P_{\pm}^{(3)}\!\mp\!\frac{2\kappa\Gamma\rho_{11}^{(0)}\Omega_p^{(\mp)}}{(\delta_p+ i\Gamma)(2\delta_B\pm i\gamma_0)}\left(\frac{|\Omega_{WM}^{(\pm)}|^2}{\delta_{4}^2}\right),\;\;\quad\\
\frac{\partial \Omega_{\pm}}{\partial z}&\!=\!&P_{\pm;WM}^{(3)}(\omega_p).
\end{eqnarray}
\end{subequations}
Here, for simplicity we have taken $\delta_4\gg\gamma_{41}=\gamma_{43}=\Gamma$. The first term in Eq. (3a) is the same single probe phase-modulation contribution as given in Eq.(1a).  The second term arises from the cross-phase modulation by the WM field which drives the cross components of the probe field through the Maxwell equations (3b).

\vskip 5pt
\noindent Equation (3a) has two important features: (1) It has the same magnetic resonance denominator as in Eq. (1a), indicating that field sensitivity is maintained (Fig. 2a); and (2) It provides a cross-component parametric WM driving source term that results in probe field propagation gain and polarization rotation enhancement (Fig. 2b). 
The central idea of the new scheme is therefore to lift the self-limiting symmetry restriction and coherently amplify the extremely weak nonlinear optical-polarization-rotation signal without introducing additional broadening effects. In addition to significantly increasing the signal/SNR of the nonlinear magneto-optical effect, the new scheme also substantially reduces the necessary initial probe field intensity, thereby enabling operation under even lower Zeeman de-coherence conditions. This latter virtue is critically important in extremely weak magnetic field sensing applications using systems with Zeeman de-coherence rates $\gamma_0<$ 1 Hz.  Indeed, for very small $\gamma_0$ the two-photon saturation effect of the strong probe required in the current state-of-art technology (due to its poor SNR) begins to detrimentally broaden the Zeeman magnetic sub-levels \cite{note3a}. We stress that the Zeeman-coherence parametric WM scheme reported here is neither a pump-probe scheme \cite{note6} nor an electromagnetically-induced transparency scheme \cite{Tslope,note3e,note4,note6a,note6b}. It is amply clear that the three-state $\Lambda-$scheme EIT process, which is not a WM process, cannot produce such large SNR enhancements because of the strong pump intensity required \cite{note6c} for transparency at the probe frequency.

\vskip 5pt
\noindent Figures 2a-2f show numerical calculations where the density matrix equations of motion and the Maxwell equations for the probe circular components are evaluated simultaneously. In Fig. 2a, the nonlinear polarization rotation of the Zeeman-coherence parametric WM scheme (red curve) is plotted together with the widely-used single-probe-beam $\Lambda-$scheme (blue curve) as a function of $\delta_B$ at $z$ = 1 cm.  In this case, the result from the single-probe-beam scheme must be multiplied by a factor of 100 in order to match the result of the WM scheme. The identical slope of the two schemes after scaling the blue curve, however, attests to the preservation of field sensitivity (see discussion on Fig. 3a), as predicted by Eqs. (1a) and (3a).  Figure 2b shows the nonlinear polarization rotation as a function of $z$ for the single-probe-beam scheme (blue curve) and the parametric WM scheme (red curve). The pronounced nonlinear growth of the polarization rotation signal (red curve) highlights the superior performance of the WM scheme. We stress that this performance cannot be matched by the single-probe-beam $\Lambda-$scheme which, by the requirement of symmetry and energy conservation, is limited to the simple linear growth behavior shown in Eq. (2). In Figures 2c-2f we show the nonlinear polarization rotation and the normalized $\Omega_p^{(+)}$ component of the probe field as a function of $z$ and $\delta_B$ for both schemes where the self-limiting effect of the single-probe-beam $\Lambda-$scheme [Eq. (2)] is clearly exhibited in Fig. 2f.

\begin{figure}
  \centering
  \includegraphics[width=8.5 cm,angle=0]{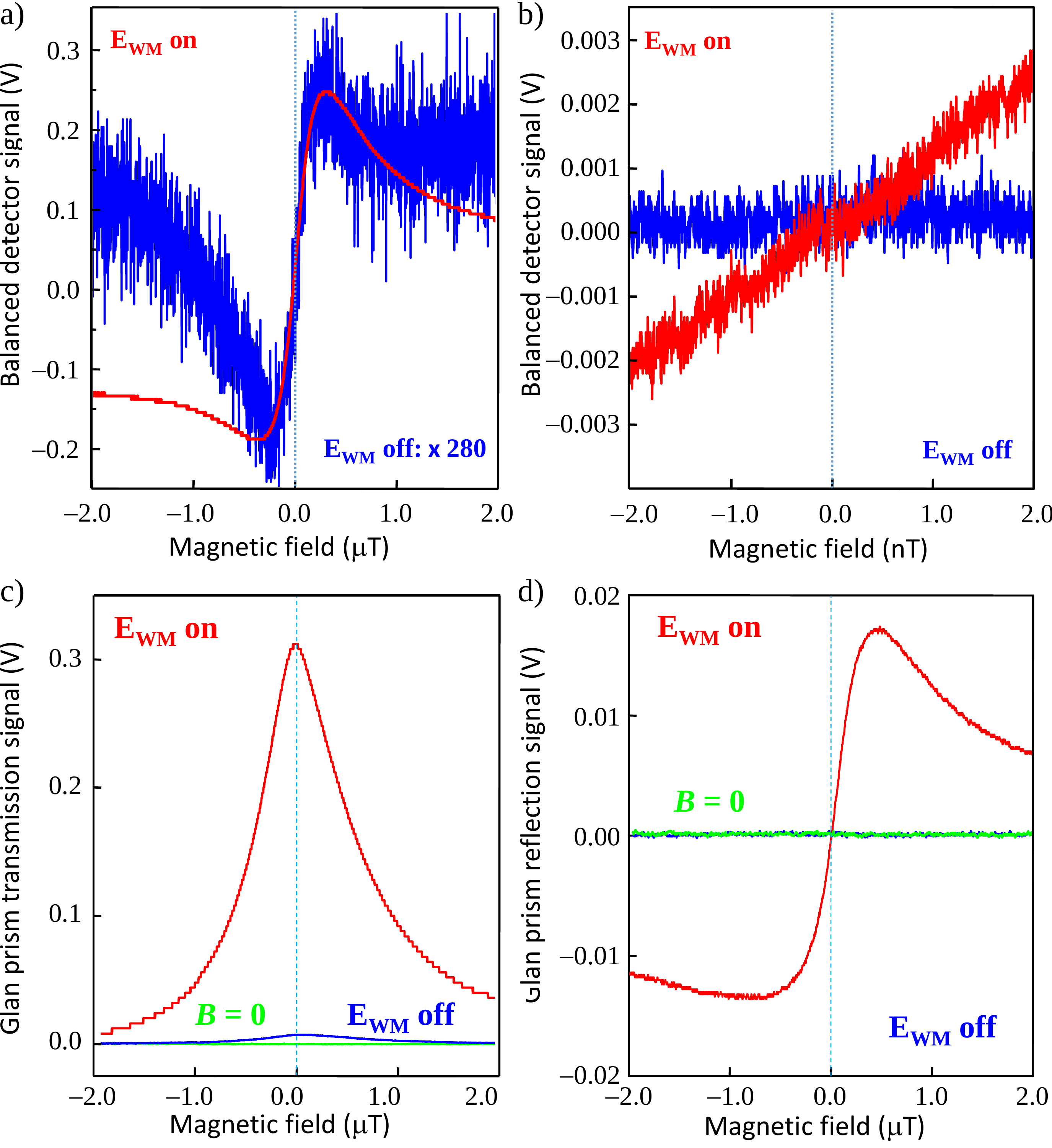}
  \caption{{\bf Optically-amplified nonlinear optical polarization rotation effect and dark resonance} (a) and (b): Balanced detector signal. (a): $I_p$ = 1.9 mW/cm$^{2}$ ($\delta_p/2\pi = -$400 MHz, $F=2\rightarrow F'=1$). $I_{WM}$ = 1.2 mW/cm$^{2}$ ($\delta_{4}/2\pi = -$120 MHz $F=2\rightarrow F{''}=3$). Data (blue) is scaled to show sensitivity preservation. The blue trace is the single-probe-beam $\Lambda-$scheme (with 128 scans averaged) and the red trace is the Zeeman-coherence parametric WM scheme (single scan without averaging).  (b): Magnetic field scan between $\pm$2 nT without slope re-scaling. (c) and (d): Signals from the orthogonal ports of a Glan prism with low-gain detectors. Temperature is 350 K. Green curves: No magnetic field. }
\end{figure}
\vskip 5pt
\noindent
Experimentally, we counter-propagate the WM light with respect to the probe light to reduce the signal background. Both linearly-polarized and circularly polarized WM light fields were studied \cite{note8}.  In the case of linearly-polarized WM light, the polarization of the light field was rotated with respect to the polarization of the probe light field to maximize SNR enhancement, forming the cross-polarization configuration of the Zeeman-coherence parametric WM scheme. Figure 3a shows a typical nonlinear magneto-optical rotation enhancement signal obtained with the parametric WM technique in our magnetic field shield-limited setup using a circularly polarized WM laser \cite{note8}.  When the energy-providing WM light field $\vec{E}_{WM}$ is turned on, we routinely observe a factor of $>$200 (Fig. 3a, red trace) SNR enhancement over the usual single-beam $\Lambda-$scheme (Fig. 3a, blue trace). The slope of the scaled magnetic resonance signal, however, remains unchanged. This indicates that the WM field does not appreciably affect magnetic field sensitivity, and this is in agreement with Eq. (3a) \cite{wang}. 
Figure 3b shows a magnetic field scan near the center of Fig. 3a without slope rescaling. 

\vskip 5pt
\noindent Figures 3c and 3d show the output of two low-gain detectors attached to the two orthogonal ports of a Glan polarizer that intercepts 50 \% of the exiting probe light. With the WM field $\vec{E}_{WM}$ turned off, a small dark resonance \cite{darkR1,darkR2,darkR3,darkR4} is observed (Fig. 3c, blue trace) in the transmission port and no signal can be detected from the orthogonal reflection port (Fig. 3d, blue trace).  When the $\vec{E}_{WM}$ field is turned on, however, a significantly WM-amplified resonance is detected in the transmission port (Fig. 3c, red trace). Correspondingly, the orthogonal reflection port registers an enormous increase in the dispersion signal (Fig. 3d, red trace), further testimony to the optical dispersion enhancement effect by the energy-transfer light. The amplified dark resonance is key evidence of the wave-mixing process. Here, the dispersion signal is enhanced by a factor of $\simeq$ 300 and can be easily detected without employing heterodyne methods, demonstrating the robustness of the novel SNR enhancement process. We note that we have observed more than three orders of magnitude optical SNR enhancement \cite{note7} with both lasers detuned as much as 4 to 8 Doppler line widths from resonance at 311 K.    

\begin{figure}
  \centering
  \includegraphics[width=8.5 cm,angle=0]{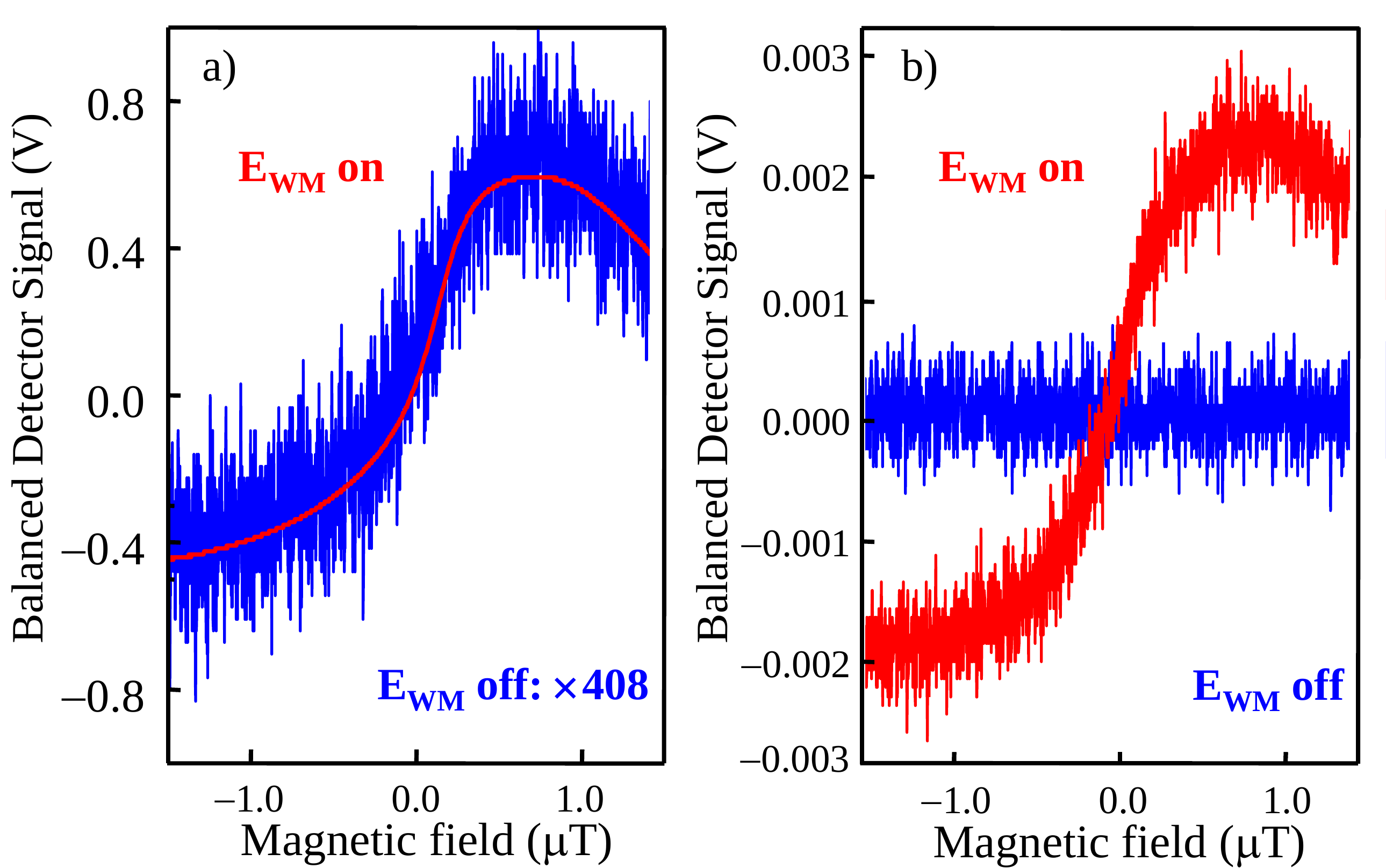}
  \caption{{\bf Optically-amplified nonlinear optical polarization rotation signal at 311 K} (a): $I_p$ = 640 $\mu$W/cm$^{2}$ with $\delta_p/2\pi = -$5 GHz ($F=2\rightarrow F'=1$). $I_{WM}$ = 80 $\mu$W/cm$^{2}$ with $\delta_{4}/2\pi = -$2 GHz ($F=2\rightarrow F^{'}=1$). Red traces: no averaging. Blue traces: average of 128 time in 3.2 seconds. (b): $I_p$ = 20 $\mu$W/cm$^{2}$ and $I_{WM}$ = 12 $\mu$W/cm$^{2}$. }
\end{figure} 
 
\vskip 5pt
\noindent
Ambient temperature magnetic field sensing at low light intensities is of great technological importance in medical research \cite{BRreview,brain,medref}. Current spin-exchange-relaxation-free based atomic magnetometers require high probe intensity, high operational temperatures \cite{Cs2}, and long signal acquisition times.  The Zeeman-coherence  parametric WM scheme demonstrated here eliminates all three of these requirements. 
These improvements taken together imply that more than a four order-of-magnitude reduction in device size is possible. This makes the technology very suitable for real-time in situ bio-magnetic field monitoring using an optical fiber. Figure 4a shows more than a 400$\times$ enhancement at human body temperatures (311 K) using a probe light intensity $I_p$ similar to those reported in literature \cite{budker1,budker1a}.  Here, the intensity of the WM light is $I_{WM}\le 0.125I_{p}$, demonstrating the impressive efficiency of the process. When the probe intensity is reduced to 20 $\mu$W/cm$^2$ (more than an order of magnitude weaker than the lowest intensity reported to date \cite{budker1}), the single-probe-beam $\Lambda-$scheme produces no detectable signal (blue trace in Fig, 4b).  The Zeeman-coherence parametric WM scheme, however, yields a clear signal with a surprisingly good SNR (red trace in Fig. 4b).  We stress that all data traces reported here were taken in less than 3.2 seconds (25 ms magnetic field sweep with 128 averaged) in an active interaction volume of $\sim$ 0.08 cm$^3$ \cite{kitching2}, exemplifying the superior performance and great potential of this new method in fast, real world applications. Indeed, such superior results have even been observed at 295 K where under the similar low power conditions the widely-used single-probe-beam method fails to produce any detectable signal, even with long data acquisition times and aggressive averaging.

\begin{figure}
  \centering
  \includegraphics[width=8.5 cm,angle=0]{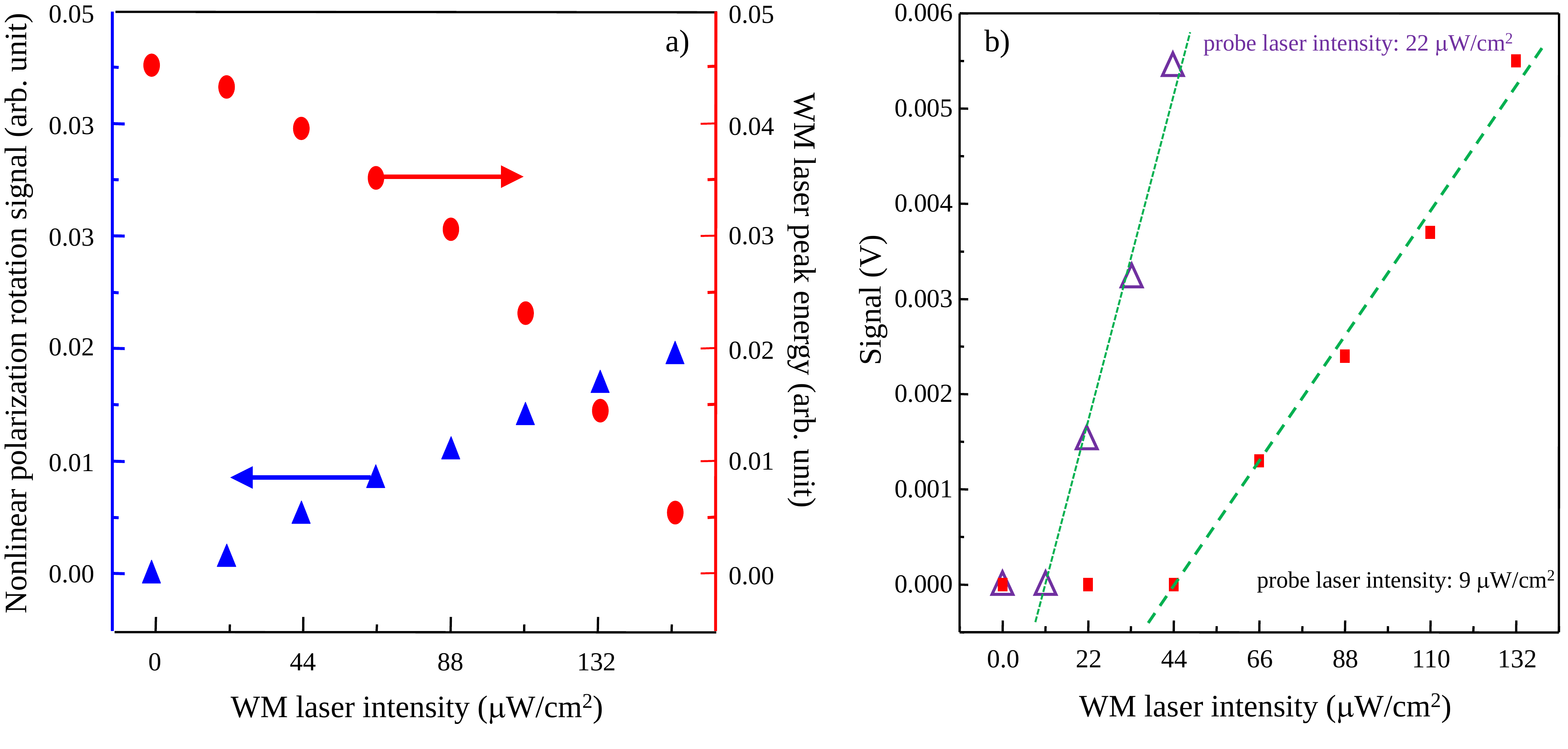}
  \caption{{\bf Zeeman-coherence cross-polarization wave mixing signature} (a): Peak-peak nonlinear polarization rotation signal (blue triangles) and the normalized WM field absorption (output WM light, red ovals) as a function of the WM light input intensity. $I_p$ = 22 $\mu$W/cm$^2$, $\delta_{4}/2\pi$ = $-$ 2 GHz, $\delta_{p}/2\pi$ = $-$ 3 GHz. (b): The fast rising of the signal as the WM field increases is the clear evidence of the wave mixing process. The lines are guides to the eye to show the presence of a ``WM-triggered-lasing-threshold" as expected from the onset of amplification/net-directional-energy-flow in the WM process. A 1-mV background has been subtracted.}
\end{figure}

\vskip 5pt
\noindent
Further evidence of coherent wave-mixing and direction energy transfer of the Zeeman-coherence cross-polarization parametric WM magnetometer is shown in Figs. 5a and 5b where the SNR enhancement is plotted against the power of the probe field and the WM field at T = 311 K. In Fig. 5a, the absorption (energy loss) of one circular component of the WM light field (red ovals) is shown as a function of the WM laser input intensity.  As the input WM laser intensity increases, the nonlinear polarization rotation signal (blue triangles) increases and the peak of the normalized WM light field component at the exit of the medium decreases rapidly, indicating significant energy flows (circulation) from the WM field to the probe field in a coherent amplification process. In Fig. 5b the peak-peak value of the signal is plotted as a function of WM power with the probe power fixed at 9 $\mu$W/cm$^2$ and 22 $\mu$W/cm$^2$, respectively.  The rapid growth of the signal with a clear threshold is indicative of a highly efficient coherent amplification (stimulated emission) process.  As expected, the growth rate becomes much more steep with a probe intensity of 22 $\mu$W/cm$^2$ (open triangles) where the WM laser has a comparable intensity.

\begin{figure}
  \centering
  \includegraphics[width=8.5 cm,angle=0]{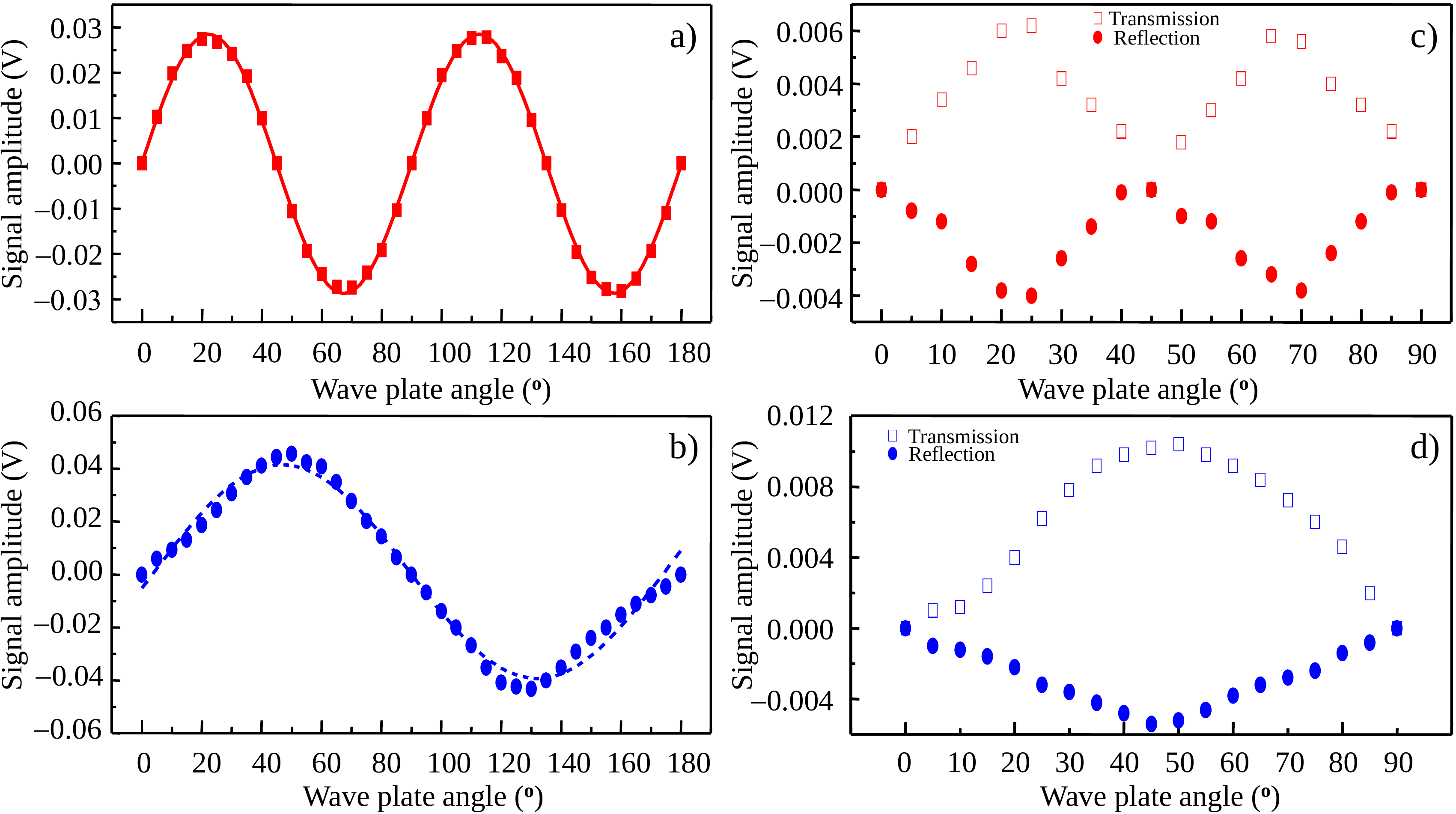}
  \caption{{\bf Cross-polarization angular-dpendence of the nonlinear polarization rotation} Balanced-detector signal (peak-peak): 1/2 wave plate (a) and 1/4 wave plate (b). Low-gain detectors from the orthogonal ports of a Glan prism: 1/2 wave plate (c) and 1/4 wave plate (d). $I_P$ = 32 $\mu$W/cm$^{2}$ and $I_{WM}$ = 20 $\mu$W/cm$^{2}$. Other parameters are the same as in Fig. 4. }
\end{figure}

\vskip 10pt
\noindent Although the cross-polarization aspect of the new scheme has some similarities with cross-polarization wave mixing in nonlinear optical crystals \cite{french1,french2}, the physics of the two processes are fundamentally different.  The underlying physics of the new scheme is Zeeman-coherence parametric WM, whereas with nonlinear optical crystals it is based on an intensity-modified nonlinear susceptibility process observable only in the ultra-high peak power regime.  The enhancement in nonlinear crystals is very weak and no intermediate state with appreciable Zeeman coherence is present. Furthermore, the ellipsometry of cross-polarization effects in crystals, created by an ultra-short pulsed probe laser field in a single-probe-beam configuration, is critically dependent upon crystal symmetry.  
In a cubic crystal with four-fold symmetry, maximum single-probe-beam cross-polarization effects have been observed when the polarization of the probe field makes angle of $\beta$ = 22.5$^{\rm o}$ and 45$^{\rm o}$ with respect to the crystal's fast axis. Interestingly, although
atomic vapors lack such crystalline symmetry properties \cite{note2}, the Zeeman-coherence parametric WM scheme demonstrated here also exhibits similar cross-polarization angular effects. In Figs. 6a-6d we plot the nonlinear polarization rotation as a function of the angle between the polarizations of the probe and WM fields. By varying this angle, maximum SNR enhancement at 22.5$^{\rm o}$ (45$^{\rm o}$) for the linearly (elliptically) polarized WM fields in an atomic vapor have been observed. Such angular dependence was not treated in the current theoretical model and is beyond the scope of this study.

\vskip 5pt
\noindent
The symmetry-breaking Zeeman-coherence parametric WM scheme demonstrated here exhibits superior SNRs in extremely weak magnetic field sensing applications. With a more sophisticated magnetic field shields the new scheme will significantly improve the sensitiveness of in extremely weak magnetic field sensing applications. We stress that the principle is directly applicable to other optical-atomic magnetometers where nonlinear optical polarization rotation is the central principle of the operation. This giant nonlinear optical polarization rotation enhancement effect may lead to breakthroughs in real-time human organ bio-magnetic field mapping. It may also lead to possible detection of the magnetic fields generated by collective ion-exchange processes in crystals and materials, providing the ability to observe crystallographic growth processes in real time as new exotic molecular structures are forming. In addition, it may lead to detection of extremely weak magnetic fields in space, nuclear magnetic resonances, the magnetism of solid-state materials and even ultra-cold atomic quantum gases such as Bose condensates and degenerate Fermion gases.




\begin{thebibliography}{1}
\bibitem{book1} A.K. Zvezdin and V.A. Kotov,  {\it Modern Magento-optics and Magneto-optical Materials}. Taylor \& Francis Group, New York (1997).


\bibitem{BRreview} D. Budker and M. Romalis, Nature Physics, {\bf 3} 227 (2007).

\bibitem{book2} D. Budker and D.F.J. Kimball, {\it Optical Magnetometry}. Cambridge University Press (2013) 

\bibitem{Tslope} V.A. Sautenkov, M.D. Lukin, C.J. Bednar, I. Novikova, E. Mikhailov, M. Fleischhauer, V.L. Velichansky, G.R. Welch, and M.O. Scully, Phys. Rev. A {\bf 62}, 023810, (2000).

\bibitem{note3e} M.O. Scully and M. Fleischhauer, 
Phys. Rev. Lett. {\bf 69}, 1360-1363 (1992).

\bibitem{note4} P. Siddons, C.S. Adams, and I.G. Hughes, Phys. Rev. A {\bf 81}, 043838 (2010).

\bibitem{shen} Y.R. Shen, {\it The Principles of Nonlinear Optics}, John Wiely \& Sons, New York, 1984.

\bibitem{happer1} W. Happer and A.C. Tam, Phys. Rev. A {\bf 16}, 1877-1891 (1977).

\bibitem{happer2} C.J. Erickson, D. Levron, W. Happer, S. Kadlecek, B. Chann, L.W. Anderson, T.G. Walker, Phys. Rev. Lett. {\bf 85}, 4237-4240 (2000).

\bibitem{walker} S. Kadlecek, L.W. Anderson, T.G. Walker, Phys. Rev. Lett. {\bf 80}, 5512-5515 (1998).

\bibitem{romalis1} J.C. Allred, R.N. Lyman, T.W.
Kornack, and M.V. Romalis, Phys. Rev. Lett. {\bf 89}, 130801 (2002).

\bibitem{romalis2} I.K. Kominis, T.W. Kornack, J.C. Allred, and M.V. Romalis, Nature {\bf 422}, 596-599 (2003).

\bibitem{romalis3} I.M. Savukov and M.V. Romalis, Phys. Rev. A {\bf 71}, 023405 (2005).

\bibitem{budker1} D. Budker, V. Yashchuk, and M. Zolotorev, Phys. Rev. Lett. {\bf 81}, 5788-5791 (1998)

\bibitem{budker1a} D. Budker, D.F. Kimball, S.M. Rochester, and V.V. Yashchuk, Phys. Rev. Lett. {\bf 83}, 1767-1770 (1999).


\bibitem{russian1} E.B. Aleksandrov et al., Optics Spectrosc. {\bf 78}, 292–298 (1995).

\bibitem{russian2} D. Budker, D.F. Kimball, S.M. Rochester, V.V. Yashchuk, and M. Zolotorev, Phys. Rev. A {\bf 62}, 043403 (2000). 

\bibitem{kitching} R. Jimenez-Martinez, S. Knappe, and J. Kitching, Rev. Sci. Instrum. {\bf 85}, 045124 (2014).



\bibitem{ref17a} B. Patton, E. Zhivun, D. C. Hovde, and
D. Budker, Phys. Rev. Lett. {\bf 113}, 013001 (2014).

\bibitem{note3} With $\delta_p/2\pi$ = 5 GHz the two-photon transition is already saturated with only $\Omega_{p}/2\pi\simeq$ 300 kHz for $\gamma_0/2\pi\le$ 5 Hz, and this introduces significant Zeeman-state broadening.  

\bibitem{lutwm1} K.J. Jiang, L. Deng, and M.G. Payne, Phys. Rev. Lett. {\bf 98}, 083604 (2007).

\bibitem{lutwm2} L. Deng, M.G. Payne, and W.R. Garrett, Opt. Commun. {\bf 242}, 641 - 647 (2004).

\bibitem{note3a} Wojciech Gawlik and Szymon Pustelny, Photonic Lett. of Poland {\bf 1}, 34 - 36 (2009).  

\bibitem{note6} Since $I_{WM}<I_{p}$ and the lasers are far-detuned from resonances the WM scheme is not the pump-probe scheme used in other magnetometer research.  

\bibitem{note6a} M. R. Blackman and B. Varcoe, Proceedings of the International Conference on Biomedical Electronics and Devices, Porto, Portugal, January 14-17, 2009.

\bibitem{note6b} V.I. Yudin, A.V. Taichenachev, Y.O. Dudin, V.L. Velichansky, A.S. Zibrov, and S.A. Zibrov Phys. Rev. {\bf 82}, 033807 (2010)

\bibitem{note6c} A.B. Matsko, O. Kocharovskaya, Y. Rostovtsev, G.R. Welch, A.S. Zibrov, and M.O. Scully, Adv. Atm. Mol. Opt. Phys. {\bf 46}, 191 (2000). 

\bibitem{note8} 
Both linearly and circularly polarized WM were studied and a circularly-polarized WM light field was found to increase the SNR by an additional 25\% due to multi-channel interference processes not shown in Fig. 1b. A circularly polarized WM field in a system without spin preparation (multiple transitions with different strengths), however, introduces small differential frequency shifts that affect the absolute zero field calibrations.  Because these shifts were below our shield limit, and for the purpose of demonstrating the largest enhancement, all data reported here except Fig. 6 were taken using a circularly polarized WM field without spin preparation. For the same reason, noise spectrum measurement was not provided (waiting for a new shield system).  

\bibitem{wang} We tested the WM scheme using a non-spin-polarized SERF magnetometer with a sensitivity of $8fT/\sqrt{Hz}$ (resolution $<$ 1 nT) and observed similar large SNR enhancements with sensitivity preservation (R.Q. Wang, to be published).

\bibitem{darkR1} A. Nagel, L. Graf, A. Naumov, E. Mariotti, V. Biancalana, D. Meschede, and R. Wynands, Europhys. Lett. {\bf 44}, 31-36 (1998).

\bibitem{darkR2} D. Budker, W. Gawlik, D.F. Kimball, S.M. Rochester, V.V. Yashchuk, and A.  Weis, Rev. of Mod. Phys. {\bf 74}, 1153-1201 (2002).


\bibitem{darkR3} M. St$\ddot{a}$ler, S. Knappe, C.  Affolderbach, W. Kemp, and R. Wynands, Europhys. Lett. {\bf 54}, 323–328 (2001). 

\bibitem{darkR4} Ch. Andreeva1, G. Bevilacqua, V.
Biancalana, S. Cartaleva1, Y. Dancheva1, T.
Karaulanov, C. Marinelli, E. Mariotti, and L. Moi, Appl. Phys. B {\bf 76}, 667 (2003)

\bibitem{note7} A factor of 2 further improvement in SNR has been observed by first optically pumping the non-probe-accessed hyperfine states in the $F=1$ grounds state manifold.

\bibitem{brain} Cort Johnson, Peter D. D. Schwindt and Michael Weisend, Appl. Phys. Lett. {\bf 97}, 243703 (2010).

\bibitem{medref} R. Wyllie, M. Kauer, R.T. Wakai, and T.G. Walker, Optics Lett. {\bf 37}, 2247-2249 (2012).

\bibitem{Cs2} C. Affolderbach, M. St$\ddot{a}$hler, S. Knappe, and R. Wynands, Appl. Phys. B 75, 605–612 (2002).

\bibitem{kitching2} S. Vishal, K. Svenja, P.D.D. Schwindt, and J. Kitching, Nature Photonics {\bf 1}, 649 - 652 (2007).

\bibitem{french1} A. Jullien, O. Albert, G. Chériaux, J. Etchepare, S. Kourtev, N. Minkovski, and S.M. Saltiel, J. Opt. Soc. Am. B {\bf 22}, 2635-2641 (2005).

\bibitem{french2} N. Minkovski, G.I. Petrov, S.M. Saltiel, O. Albert, and J. Etchepare, J. Opt. Soc. Am. B {\bf 21}, 1659-1664 (2004).

\bibitem{note2} A. Weis, J. Wurster, and S.I. Kanorsky,  J. Opt. Soc. Am. B {\bf 10}, 716-724 (1993).









\end{thebibliography}
\end{document}